\documentclass[sn-mathphys,Numbered]{sn-jnl}

\usepackage{graphicx}%
\usepackage{multirow}%
\usepackage{amsmath,amssymb,amsfonts}%
\usepackage{amsthm}%
\usepackage{mathrsfs}%
\usepackage[title]{appendix}%
\usepackage{xcolor}%
\usepackage{textcomp}%
\usepackage{manyfoot}%
\usepackage{booktabs}%
\usepackage{algorithm}%
\usepackage{algorithmicx}%
\usepackage{algpseudocode}%
\usepackage{listings}%
\usepackage{subfig}%
\usepackage{multirow}%
\usepackage{enumitem}%
\usepackage{microtype}%
\usepackage{xspace}
\usepackage{caption}
\usepackage{lmodern}
\usepackage{anyfontsize}
\usepackage{hyperref}
\usepackage{breakurl}

\newcommand\PC{\textsc{Math\-Check}\xspace}
\newcommand{\proofsizetwo}{6.1\xspace} 
\newcommand{\proofsizethree}{40.3\xspace} 
\newcommand\AMS{\textsc{Alpha\-MapleSAT}\xspace}

\newtheorem{theorem}{Theorem}
\newtheorem{observation}[theorem]{Observation}
\newtheorem{definition}{Definition}%

\begin{document}

\title[A SAT+CAS Attack on the Minimum Kochen--Specker Problem]{A SAT Solver + Computer Algebra Attack on the Minimum Kochen--Specker Problem}


\author[1]{\fnm{Zhengyu} \sur{Li}}\email{brian.li@gatech.edu}

\author[2]{\fnm{Curtis} \sur{Bright}}\email{cbright@uwindsor.ca}

\author[1]{\fnm{Vijay} \sur{Ganesh}}\email{vganesh45@gatech.edu}

\affil[1]{School of Computer Science, Georgia Institute of Technology}

\affil[2]{School of Computer Science, University of Windsor}


\abstract{One of the fundamental results in quantum foundations is the Kochen–Specker (KS) theorem, which states that any theory whose predictions agree with quantum mechanics must be \emph{contextual}, i.e., a quantum observation cannot be understood as revealing a pre-existing value.
The theorem hinges on the existence of a mathematical object called a KS vector system. While many KS vector systems are known, the problem of finding the minimum KS vector system in three dimensions (3D) has remained stubbornly open for over 55 years.

To address the minimum KS problem, we present a new verifiable proof-producing method based on a combination of a Boolean satisfiability (SAT) solver and a computer algebra system (CAS) that uses an {\it isomorph-free orderly generation} technique that is very effective in pruning away large parts of the search space. Our method shows that a KS system in 3D must contain at least 24 vectors. We show that our sequential and parallel Cube-and-Conquer (CnC) SAT+CAS methods are significantly faster than SAT-only, CAS-only, and a prior CAS-based method of Uijlen and Westerbaan. Further, while our parallel pipeline is somewhat slower than the parallel CnC version of the recently introduced Satisfiability Modulo Theories (SMS) method, this is in part due to the overhead of proof generation. Finally, we provide the first computer-verifiable proof certificate of a lower bound to the KS problem with a size of \proofsizethree TiB in order 23.}


\maketitle

\section{Introduction}\label{introduction}
Quantum Mechanics (QM) is often described as one of the most successful physical theories of all time, and yet many questions regarding the very foundations of QM remain unresolved. To address these foundational issues, many interpretations of QM (i.e., mappings from mathematical formalisms of QM to physical phenomena) have been proposed. Hidden-variable theories are attempts at understanding counterintuitive QM phenomena through a deterministic lens by positing the existence of (possibly) unobservable physical entities or hidden variables~\cite{sep-kochen-specker} that standard QM theory does not account for (and hence is deemed incomplete). Over the years, many constraints have been imposed on hidden-variable theories, e.g., Bell's inequalities that rule out the possibility of \emph{local} hidden-variable theories that are also in agreement with the predictions of QM~\cite{bell2004speakable}. In a similar vein, Simon Kochen and Ernst Specker~\cite{Kochen1967-KOCTPO-3} proved their famous Kochen--Specker (KS) theorem in 1967 (and independently by John Bell in 1966~\cite{RevModPhys.38.447}) that essentially asserts that non-contextual hidden variable theories cannot reproduce the empirical predictions of QM.

The KS theorem rules out non-contextual hidden-variable theories via the existence of a finite set of three-dimensional vectors, referred to as a \emph{KS vector system}~\cite{Kochen1967-KOCTPO-3}. A KS vector system (or simply a KS system) is a combinatorial object that witnesses a contradiction between non-contextuality (i.e., the assumption that observables can be assigned values prior to measurement and independent of measurement context) and the SPIN axiom of QM\@. The first KS vector system, discovered in 1967, contains 117 vectors~\cite{Kochen1967-KOCTPO-3}. Another theorem that relies on the existence of KS systems in an essential way is the ``Free Will'' theorem of John Conway and Simon Kochen~\cite{conway2006free}.  

Since the publication of Kochen and Specker's theorem in 1967, physicists and mathematicians have wondered about the cardinality of the smallest-sized KS vector system (see Table~\ref{tbl:history} and Section~\ref{sec:related}). Finding the minimum KS system, referred to as the minimum KS problem, is not only of scientific and historical interest but also has direct applications in quantum information processing~\cite{Caas2014}.  For example, finding a minimum KS system could enable applications in the security of quantum cryptographic protocols based on complementarity~\cite{cabello2011hybrid}, zero-error classical communication~\cite{cubitt2010improving}, and dimension witnessing~\cite{guhne2014bounding}. Further, the large size of all known KS systems has hindered physicists from using them for empirical tests of the KS theorem, similar to the empirical tests of Bell's theorem~\cite{santos1992critical}. 

\subsection{Two definitions of the Kochen-Specker System} \label{def}

There are two definitions of the KS system widely used in literature.
The ``original'' KS set definition used in this paper (Section \ref{background}) contains only the
vectors necessary to prove the KS theorem mathematically.
This ``original'' definition of a KS set is the one originally
used by Kochen and Specker themselves. KS systems developed by this definition are commonly referred to as ``original KS systems'' \cite{larsson2002kochen}.
However, from an experimental perspective, another definition that requires additional vectors in the KS system is used, since constructing
the set in practice would involve vectors not explicitly
needed in the mathematical proof. Specifically, this definition requires that every pair of vectors in a 3-dimensional KS set belongs to a set of 3 mutually orthogonal vectors.  KS systems developed by this alternative definition are commonly referred to as ``extended KS systems'' \cite{larsson2002kochen}.

Both definitions are well-known and used extensively in the literature. For examples, the `original' definition used in this paper is also used in \cite{Kochen1975}, \cite{peres1991two}, \cite{peres1997quantum},
\cite{bub1996schutte}, \cite{conway2006free}, \cite{Cabello2006}, \cite{arends2011searching},
\cite{uijlen2016kochen}, \cite{KirchwegerPeitlSzeider23}, while the other definition is used in \cite{larsson2002kochen}, \cite{pavivcic2017arbitrarily}, \cite{budroni2022kochen}, \cite{held2009kochen}, \cite{pavivcic2019hypergraph}, \cite{pavivcic2005kochen}, \cite{pavicic2023quantum}. As a result of the difference in the two definitions, the lower bound on the original KS system and the extended KS system are also different, as shown in Table \ref{tbl:def}.

John Conway has stressed the problem of finding the minimum number
of three-dimensional vectors necessary to prove the ``Free Will theorem''
in public lectures on the topic (see \cite{arends2011searching}). Thus, knowing the smallest `original' KS set is of interest since
such a set would correspond to a proof of the Free Will theorem
using the fewest number of three-dimensional directions.  This
in a certain sense would lead to the `simplest' proof of the theorem.
We believe the question of the minimal size of an `original'
KS set is theoretically interesting (independent of what
the minimal size of an `experimental' KS set is). In this paper, we investigate the lower bound of the original KS system. However, the paradigm proposed in this paper is easily adaptable and scalable, and the approach
can be applied to both definitions given the appropriate SAT encoding of the problem.

\begin{table}
\begin{minipage}{200pt}
\begin{tabular}{lll}
Discoverers & Original System & Extended System \\ \hline
Kochen, Specker \cite{Kochen1967-KOCTPO-3} & 117 & 192 \\
Sch\"{u}tte \cite{bub1996schutte} & 33 & 49 \\
Peres \cite{Peres1991} & 33 & 57 \\
Conway, Kochen \cite{peres1997quantum}  & 31 & 51 \\

\end{tabular}
\end{minipage}
\caption{Size of 3-dimensional KS systems and how they differ based on the definition used, as discussed in Subsection \ref{def}. In addition, Pavi{\v{c}}i{\'{c}} and Megill \cite{Pavicic:2022ezy} discovered many other unique 3-dimensional extended KS systems with 51, 53, 54, 55, 57, 69, etc. vectors through automated generation.}\label{tbl:def}%
\end{table}

\subsection{The SAT+CAS Paradigm for Hard Combinatorial Problems} \label{sat}

In recent years we have witnessed the dramatic impact of satisfiability (SAT) solvers---computer programs that take as input Boolean logic formulas and decide whether they have solutions---in areas as diverse as AI, software engineering, program verification, program synthesis, and computer security~\cite{biere2021handbook,ganesh2020unreasonable}. Unfortunately, despite these fantastic achievements of SAT solvers, they struggle with certain problems such as those containing many symmetries~\cite{Metin2018} or those requiring the usage of more advanced mathematical theories than propositional logic~\cite{bright2022science}. Much work has been done to remedy these drawbacks, including the development of sophisticated symmetry breaking techniques~\cite{10.5555/1630659.1630699} and the development of solvers that support richer logic such as ``SAT modulo theories'' or SMT solvers~\cite{BarFT-SMTLIB}.  However, the mathematical support of SMT solvers is quite limited when compared with the vast mathematical functionality available in a modern computer algebra system (CAS).

In response to this need for a solver that combines the efficient search capabilities of SAT solvers with the mathematical knowledge available in CASs, a new kind of solving methodology was developed in 2015 by Zulkoski, Ganesh, and Czarnecki~\cite{ZulkoskiGC15} and independently by \'Abrah\'am~\cite{abraham2015building}. This SAT+CAS solving methodology has been successfully applied to many diverse problems, including circuit verification~\cite{Kaufmann2023,dfki12267}, automatic debugging~\cite{Mahzoon2018}, finding circuits for matrix multiplication~\cite{Heule2021}, computing directed Ramsey numbers~\cite{Neiman2022}, and verifying mathematical conjectures~\cite{bright2016mathcheck}. For other work in the intersection of symbolic computation and satisfiability checking, see Matthew England's summary~\cite{England2022a} of the SC-Square project. In short, the SAT+CAS methodology has found wide application in diverse fields that somehow require solving hard combinatorial problems.

\begin{table}
\begin{minipage}{200pt}
\begin{tabular}{lll}
Discoverers & Year  & Bound \\ \hline
Kochen, Specker \cite{Kochen1967-KOCTPO-3}  & 1967   & $\leq 117$   \\
Jost \cite{jost1976measures} & 1976 & $\leq 109$ \\
Conway, Kochen \cite{peres1997quantum} & 1990  & $\leq 31$ \\
Arends, Ouaknine, Wampler \cite{arends2011searching} & 2009 & $\geq 18$ \\
Uijlen, Westerbaan \cite{uijlen2016kochen}  & 2016   & $\geq 22$ \\
Li, Bright, Ganesh \cite{DBLP:conf/scsquare/LiBG22} & 2022 & $\geq23$ \\
Li, Bright, Ganesh \cite{brianaaai} /  & \multirow{2}{*}{2023} & \multirow{2}{*}{$\geq 24$} \\
Kirchweger, Peitl, Szeider \cite{KirchwegerPeitlSzeider23}
\end{tabular}
\end{minipage}
\caption{A chronology of the bounds on the size of the minimum KS vector system in three dimensions. This table should not be regarded as a comprehensive catalog of all three-dimensional KS systems (Section \ref{sec:related}); rather, it is a chronological overview highlighting the advancements in reducing the size of the minimal 3-D KS system in line with its initial definition.
The present work (presented at CanaDAM 2023)
was performed independently of Kirchweger, Peitl, Szeider (presented at IJCAI 2023).}\label{tbl:history}%
\end{table}

\begin{figure}[!ht]
\centering
\includegraphics[scale=0.7]{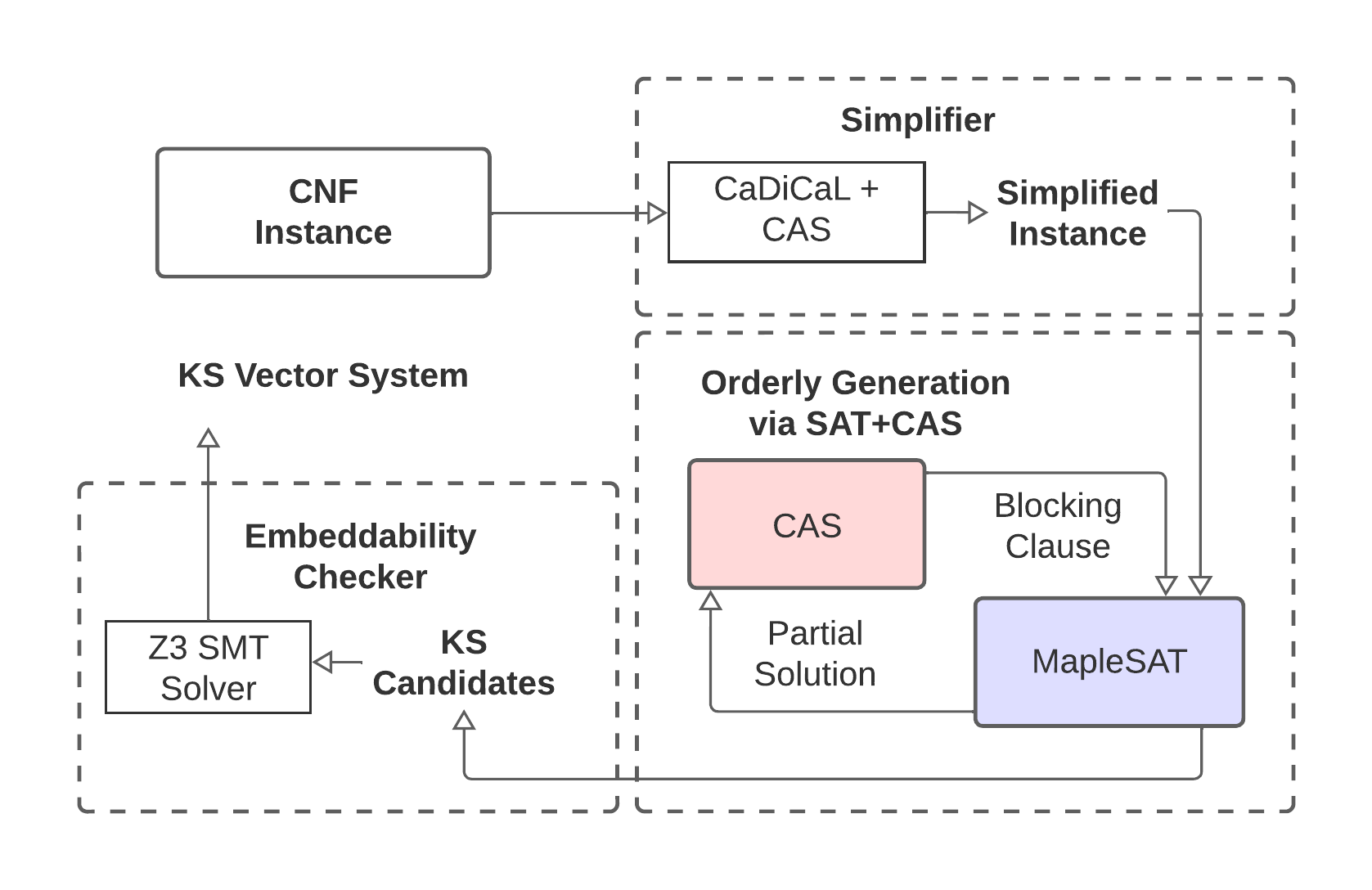}
\caption{A flowchart of our SAT+CAS based tool \PC~for solving the KS problem in the sequential setting.
The CNF instance encoding the KS problem (see Section~\ref{sec:encoding}) is simplified using CaDiCaL+CAS. The simplified instance is passed to the MapleSAT+CAS tool (see Section~\ref{orderly}) either sequentially or in parallel using Cube-and-Conquer (CnC).  Finally, an embeddability checker applies the SMT solver Z3 to determine whether the candidates are embeddable (see Section~\ref{sec:embeddability}).}\label{fig:pipeline}
\end{figure}

In this paper, we use the SAT+CAS solving methodology (see Figure~\ref{fig:pipeline}) to dramatically improve the performance of the search for KS systems compared to all previous approaches developed to prove lower bounds for the minimum KS problem (see Section~\ref{contribution}). This is made possible via a combination of the powerful search and learning algorithms used in modern SAT solvers with an ``isomorph-free exhaustive generation'' approach that prevents the duplicate exploration of isomorphic parts of the search space by the solver.
For example, such an approach was recently used to resolve the Lam's problem from projective geometry~\cite{bright2021sat}.
Although isomorph-free exhaustive generation has been used extensively in combinatorial enumeration, it has only recently been combined with SAT solving~\cite{Junttila2020,Savela2020}.

The traditional approach to preventing a SAT solver from repeatedly exploring isomorphic parts of a search space is via the use of \emph{symmetry breaking} techniques~\cite{Metin2018}.
One such symmetry breaking approach is to add ``static'' constraints to the input formula at the beginning of the search aimed at reducing the size of the search space~\cite{crawford1996symmetry,Heule2019}. Unfortunately, such an approach can be quite expensive in the sense that the number of added constraints can be large (e.g., exponential in the number of variables of the formula that encodes the problem-at-hand). Another approach is to ``dynamically''
break symmetries during the solver's search~\cite{DBLP:conf/ijcai/SellmannH05,Metin2018} such as in the SAT modulo symmetries (SMS) paradigm~\cite{kirchweger2021,KirchwegerScheucherSzeider22}.  Our approach is similar in that it also dynamically adds constraints to the problem during the solving process. However, an important difference is that the SAT+CAS paradigm is more general since it goes beyond breaking symmetries. For example, in the resolution of the smallest counterexample of the Williamson conjecture, we used the Discrete Fourier Transform (DFT) as part of the CAS computations~\cite{bright2020applying}.

\subsection{Automated Verification of Results}
Verification is of utmost importance in the context of computer-assisted proofs, given the mathematical nature of such computations---especially for nonexistence proofs.
Fortunately, the SAT+CAS paradigm naturally lends itself to automated verification, given the fact that all modern SAT solvers produce verifiable proofs.
By contrast, all previous computer-assisted proofs of lower bounds for the minimum KS problem are not verifiable.

Since our problem requires the solver to perform an exhaustive search, the validity of our nonexistence result is crucially dependent on the encodings
and the computational tools that we use.
For example, our nonexistence result crucially relies on the correctness of the SAT solver's search and the computer algebra system's isomorph-free exhaustive generation routine.
Fortunately, our SAT+CAS method generates verifiable certificates that allow an independent third party to certify that the SAT solver's search
is exhaustive and also that the facts provided by the isomorph-free generation are correct.
Thus, one does not need to trust either the SAT solver or the CAS to trust that our results are correct---instead, one only needs to trust the correctness of the proof verifier.
This is quite significant, as SAT solvers and CASs are complicated pieces of software that typically cannot be guaranteed to
be bug-free.
By contrast, a proof verifier is a much simpler piece of software that can be formally checked.
In Section~\ref{robust}, we provide details on the verification techniques that we used to certify our results.

\subsection{Our Contributions}\label{contribution} 

\begin{itemize}

\item {\bf Proof-producing SAT+CAS with Orderly Generation (OG) Method:} In this paper, we present the design and implementation of a verifiable proof-producing SAT+CAS system with orderly generation (OG) aimed at combinatorial problems (as part of the SAT+CAS tool, \PC\footnote{Code at \url{https://github.com/BrianLi009/MathCheck}}), thus showing that the minimum KS system must contain at least 24 vectors.
Also, we extend our work to complex vectors in three dimensions (3D), and thus establish a lower bound of 24 for both the real and complex KS problem.

\item {\bf Speedup over Competing Methods:} We show that our sequential and parallel Cube-and-Conquer (CnC) SAT+CAS methods are significantly faster than SAT-only, CAS-only, and prior CAS-based methods by~\cite{uijlen2016kochen}. Further, while our pipeline is somewhat slower than the recently introduced SMS method, this is in part due to the generation of verifiable proofs by our methods which added some additional complexity and slowed down our method.

\item {\bf Formal Verification of Results:} Finally, our approach provides a formal verification of the lower bound of 24 for the minimum KS problem in 3D by verifying all certificates computed by the SAT+CAS solvers in all orders up to and including order 23 (see Section~\ref{robust}). By contrast, \cite{KirchwegerPeitlSzeider23} describe a method to verify their result, but they only report having verified 5\% of the proofs in order 23.
\end{itemize}

\section{Background}\label{background}
In this section, we introduce several fundamental concepts from quantum foundations such as the SPIN axiom, 010-colorability, the KS theorem, and the KS vector system. For a deeper dive, we refer the reader to the QM section in the Stanford Encyclopedia of Philosophy~\cite{sep-kochen-specker}. We assume that the reader is familiar with Boolean logic and SAT solvers. While we provide a very brief overview of cube-and-conquer SAT solvers, we refer the reader to the Handbook of Satisfiability~\cite{biere2021handbook} for a comprehensive overview.

\subsection{The KS Theorem}

\begin{figure}[!ht]
    \centering
    \includegraphics[scale=0.4]{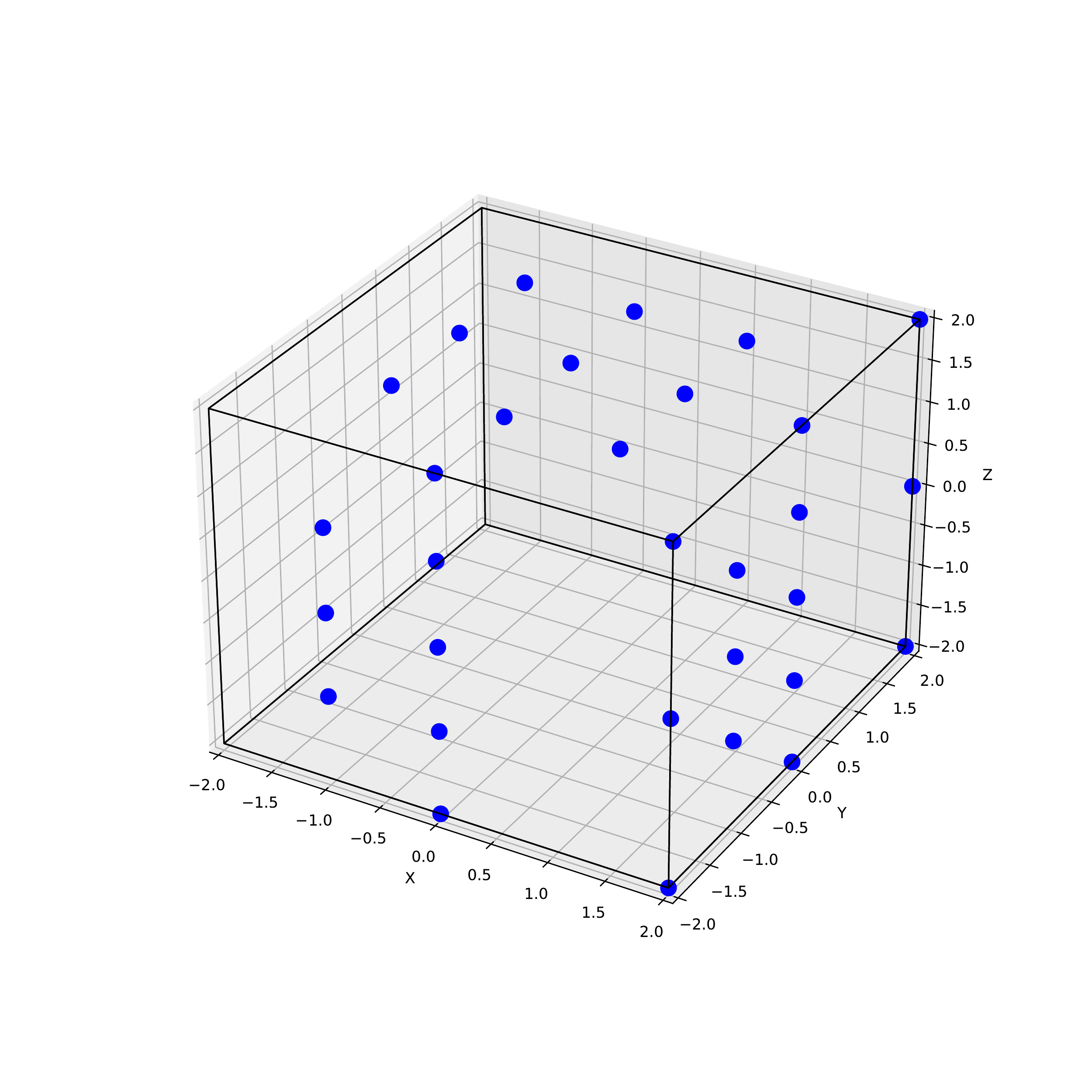}
    \caption{The 31 vectors of the smallest known KS system in three dimensions (discovered by John Conway and Simon Kochen circa 1990).
    For simplicity, the vectors have been scaled to lie on the cube with vertices $(\pm2,\pm2,\pm2)$ instead of the unit sphere.}
    \label{fig:k31}
\end{figure}

Informally, the KS theorem states that there is a contradiction between the SPIN axiom of standard QM and the assumption of non-contextuality. The Stanford Encyclopedia of Philosophy provides a comprehensive background to the KS theorem and stresses its importance in the foundations of QM~\cite{sep-kochen-specker}. The proof of the KS theorem crucially relies on the existence of a KS vector system (see Figure~\ref{fig:k31}). More precisely, exhibiting the existence of a KS vector system proves the KS theorem, which essentially states that the unit sphere is not 010-colorable (defined below).

\noindent \textbf{Spin of an Elementary Particle:} Spin is an intrinsic form of angular momentum of elementary particles whose existence can be inferred from the Stern--Gerlach experiment~\cite{GerlchDerEN}. In our context, a spin-1 particle is shot through a magnetic field in a given direction and continues undisturbed, deflects up, or deflects down---corresponding to 3 possible angular momentum states, namely $0$, $1$, and $-1$. Thus, the square of this measurement is~$0$ or~$1$. 

\vspace{0.1cm}
\noindent \textbf{SPIN axiom:} The SPIN axiom of QM states that the squared spin components of a spin-1 particle are 1, 0, 1 in three pairwise orthogonal directions of measurement.
Thus, the observable corresponding to the question ``is the squared spin $0$?''\ measured in three mutually orthogonal direction always produces \emph{yes} in exactly one direction and \emph{no} in
the other two orthogonal directions. 
We use the dual of the above form in the present work, i.e., the `010' convention rather than `101', following \citeauthor{uijlen2016kochen}.
The SPIN axiom follows from the postulates of QM and is experimentally verifiable~\cite{huang2003experimental}.

\vspace{0.1cm}
\noindent \textbf{KS Vector System:} A KS vector system can be represented in multiple ways and we describe it as a finite set of points on a sphere. As a consequence of the SPIN axiom, the squared-spin measurements along opposite directions must yield the same outcome. Therefore, two collinear vectors are considered to be equivalent. To define a KS vector system, we first formally define a vector system and the notion of 010-colorability.  For the purposes of this paper, we limit ourselves to the 3D version of the KS problem as the size of the minimum KS system
in higher dimensions is already known~\cite{Pavii2005}. 

\begin{definition}[{\bf Vector System}]
A \textbf{vector system} is a finite set of non-collinear points on the unit sphere in\/ $\mathbf{R}^3$.
\end{definition}

A $\{0,1\}$-coloring of a vector system is an assignment of 0 and~1 to each vector in the system.
The colorings of interest to us are described in the following definition.

\begin{definition}[{\bf 010-Colorability of Vector Systems}]
A vector system is \textbf{010-colorable} if there exists an assignment of 0 and~1 to each point such that:
\begin{enumerate}
    \item No two orthogonal points are assigned 1.
    \item Three mutually orthogonal points are not all assigned 0.
\end{enumerate}
\end{definition}

\begin{definition}[{\bf KS Vector System}]
A \textbf{KS vector system} is one that is not 010-colorable.
\end{definition}

\begin{definition}[{\bf Orthogonality Graph}]
For a vector system $\mathcal{K}$, define its \textbf{orthogonality graph} $G_\mathcal{K} = ( V, E )$, where $V = \mathcal{K}$, $E = \{\,(v_1, v_2): v_1, v_2 \in
\mathcal{K} \text{ and } v_1 \cdot v_2 = 0 \,\}$. 
\end{definition}

Essentially, the vertices of $G_{\mathcal{K}}$ are the vectors in $\mathcal{K}$, and there is an edge between two vertices exactly when their corresponding vectors
are orthogonal. Similarly, the notion of 010-colorability can be translated from a vector system to an orthogonality graph.

\begin{definition}[{\bf 010-colorability of Graphs}]
A graph $G$ is \textbf{010-colorable} if there is a $\{0, 1\}$-coloring of the vertices such that the following conditions are satisfied simultaneously:
\begin{enumerate}
    \item No two adjacent vertices are colored 1.
    \item For each triangle, the vertices are not all colored 0.
\end{enumerate}
\end{definition}

It is not always the case that an arbitrary graph has a corresponding vector system,
but if one does exist then we say that such a graph is \emph{embeddable}.

\vspace{0.1cm}
\begin{definition}[{\bf Embeddable Graph}]\label{embeddable}
A graph $G = (V, E) $ is \textbf{embeddable} if it is a subgraph of an orthogonality graph for some vector system.
\end{definition}

Being embeddable implies the existence of a vector system~$\mathcal{K}$ whose vectors have a one-to-one correspondence with the vertices of $G$, such that adjacent vertices are assigned to orthogonal vectors. An example of an unembeddable graph is the cyclic graph $C_4$ on 4 vertices, as the orthogonality constraints force a pair of opposite vertices to be mapped to collinear vectors (which are not allowed in this context).

\vspace{0.1cm}
\begin{definition}[{\bf KS Graph}]
An embeddable and non-010-colorable graph is called a \textbf{KS graph}.
\end{definition}

\vspace{0.1cm}
\begin{observation}
There exists a KS vector system if and only if there exists a KS graph.    
\end{observation}

\vspace{0.1cm}
\begin{definition}[{\bf KS Candidates}]
A \textbf{KS Candidate} is a satisfying assignment generated by the SAT+CAS solver.
\end{definition}

If a KS candidate is also embeddable, then it is a KS graph. As described in Section~\ref{sec:embeddability}, our solver dynamically blocks all minimal unembeddable subgraphs up to order 12, and hence our KS candidates are graphs that do not contain any unembeddable subgraphs of order up to 12.

\subsection{The Minimum KS Problem}

The minimum KS problem is to find a KS vector system of minimum cardinality, that is, a system with the fewest number of vectors in three-dimensional space (or equivalently a KS graph with the fewest number of vertices).  Every KS system has an associated KS graph, so if a KS graph with cardinality~$n$ does not exist then a lower bound on the minimum KS problem is at least $n+1$.

\subsection{Cube-and-conquer}\label{cnc}
The cube-and-conquer SAT solving paradigm was
developed in~\cite{heule2011cube} to solve hard combinatorial problems. The method applies two (possibly) different types of SAT solvers in two stages: First, a ``cubing solver'' splits a SAT instance into a large number of distinct subproblems specified by cubes---formulas of the form $x_1 \land \dots \land x_n$ where~$x_i$
are literals. Second, a ``conquering solver'' solves each subproblem under the assumption that its associated cube is true (more precisely, the conjunction of the original instance and the cube). 

The cube-and-conquer method has empirically been shown to be effective at quickly solving large satisfiability problems when the cubing solver generates many cubes encoding subproblems of similar difficulty. It has since been applied to solve huge combinatorial problems such as the Boolean Pythagorean triples problem~\cite{10.1007/978-3-319-40970-2_15}, the computation of Schur number five~\cite{heule2018schur}, and a SAT-based resolution of Lam's problem~\cite{bright2021sat}.

\section{Previous Work}\label{sec:related}

Over the last $55+$ years, many mathematicians and physicists such as Roger Penrose, Asher Peres, and John Conway have attempted to find a minimum 3-dimensional KS system (see Table~\ref{tbl:history}).
The first KS system was constructed in 1967 and it contained 117 vectors~\cite{Kochen1967-KOCTPO-3}.
A KS system with 109 vectors was found by Res Jost~\cite{jost1976measures} in 1976. Peres found a KS system of size 33 in 1991, and Sch\"{u}tte found a KS system of size 33 in 1996.
The current smallest known KS system in three dimensions contains 31 vectors and was discovered by John Conway and Simon Kochen circa 1990 (see Figure~\ref{fig:k31}).
All these discoveries were made analytically, without the assistance of computational methods. Recently, Pavi{\v{c}}i{\'{c}} and Megill \cite{Pavicic:2022ezy} applied an automated generation approach to robustly generate KS systems in odd dimensions.  This approach led to the discovery of many more three-dimensional KS systems.

In 2011, Arends, Ouaknine, and Wampler proved several interesting properties of KS graphs and leveraged them to computationally establish that a KS system must contain at least 18 vectors~\cite{arends2011searching}.
Seven years later, Uijlen and Westerbaan showed that a KS system must have at least 22 vectors~\cite{uijlen2016kochen}.
This computational effort used around 300 CPU cores for three months and relied on the \emph{nauty} software package~\cite{MCKAY201494} to exhaustively search for KS graphs.
Pavi{\v{c}}i{\'{c}}, Merlet, McKay, and Megill~\cite{Pavii2005} have improved a variation of the KS problem, one in which each vector is part of a mutually orthogonal triple (or a mutually orthogonal $d$-tuple in $d$ dimensions).
Under this restriction, they show a KS system must have at least 30 vectors in $d=3$ dimensions, and in $d\geq 4$ dimensions the minimum KS system has 18 vectors.
However, in three dimensions the gap between the lower and upper bounds of a KS system remains significant and the minimum size remains unknown.

Another way of measuring the size of a $d$-dimensional KS system is the number of mutually orthogonal ``contexts'' (cliques of size~$d$ in the orthogonality graph).
Lison\v{e}k, Badzi\c{a}g, Portillo, and Cabello~\cite{Lisonk2014} found a six-dimensional KS system with seven contexts and
showed this is the simplest possible KS system allowing a symmetry parity proof of the KS theorem.
This KS system was later experimentally used by Ca\~{n}as~et al.~\cite{Caas2014} to perform measurements
verified to arise from a quantum system rather than a classical system.

Preliminary versions of the present work were announced at the \emph{2022 SC-Square workshop}, as well as at the \emph{2023 Southeastern International Conference on Combinatorics, Graph Theory and Computing}, and at \emph{CanaDAM 2023}.
At the former two venues, we presented searches for KS systems with up to 22 vectors, and at CanaDAM 2023 we presented our work that extends this to a search for KS systems with up to 23 vectors. In each case the searches were exhaustive and no KS systems were found. Thus, a KS system in three dimensions must contain at least 24 vectors.

Kirchweger, Peitl, and Szeider~\cite{KirchwegerPeitlSzeider23}
completed an independent search for KS~systems establishing a lower bound of 24 vectors with a similar approach
as SAT+CAS but
with an SMS solver and an alternate definition of canonicity. Also, they do not use OG, as their definition of canonical does not satisfy property~(2) from Sec.~\ref{orderly}, but otherwise the SMS approach is similar in that it combines a SAT solver with a canonical checking routine. Their approach can also be used to generate proof certificates, though the certificate verification was not performed with the exception of 5\% of the certificates in the order~23 search. 
By constrast, ours formally verifies all results.

\section{SAT Encoding of the Minimum KS Problem}\label{sec:encoding}

As stated earlier, every KS vector system $\mathcal{K}$ can be converted into a KS graph $G_{\mathcal{K}}$.
Each vector in $\mathcal{K}$ is assigned to a vertex in $G_{\mathcal{K}}$, so that if two vectors are orthogonal, then their corresponding vertices are connected.

We say a KS graph is minimal if the only subgraph that is also a KS graph is itself. Arends, Ouaknine, and Wampler~\cite{arends2011searching} proved that a three-dimensional minimal KS graph must satisfy the following properties: 

\begin{enumerate}
    \item The graph does not contain the 4-cycle graph $C_4$ as a subgraph.
    \item Each vertex of the graph has a minimum degree $3$.
    \item Every vertex is part of a 3-cycle triangle graph $C_3$.
\end{enumerate}
We encode these three properties and the non-010-colorability of the KS graph in conjunctive normal form (CNF), as described below. If a SAT solver produces solutions for such an encoding, then these solutions are equivalent to graphs that satisfy all of the above-mentioned four constraints.

A simple undirected graph of order~$n$ has $\binom{n}{2}$ potential edges, and we represent each edge as a Boolean variable.  The edge variable $e_{ij}$ is true exactly when the vertices $i$ and
$j$ are connected, where $1\leq i<j\leq n$.
For convenience, we let both $e_{ij}$ and $e_{ji}$ denote the same variable since the graphs we consider are undirected.
We also use the $\binom{n}{3}$ triangle variables $t_{ijk}$ denoting that distinct vertices $i$, $j$, and $k$ are mutually connected.
In Boolean logic this is expressed as $t_{ijk}\leftrightarrow(e_{ij}\land e_{ik}\land e_{jk})$ which in conjunctive normal
form is expressed via the four clauses
$\lnot t_{ijk}\lor e_{ij}$, $\lnot t_{ijk}\lor e_{ik}$, $\lnot t_{ijk}\lor e_{jk}$, and $\lnot e_{ij}\lor \lnot e_{ik}\lor \lnot e_{jk}\lor t_{ijk}$.
Again, the indices $i$, $j$, and $k$ of the variable $t_{ijk}$ may be reordered arbitrarily for notational convenience.

\subsection{Encoding the Squarefree Constraint}
To encode the property that a KS graph must be squarefree, we construct encodings that prevent the existence of any squares in the graph. Observe that three squares can be formed on four vertices. Therefore, for every choice of four vertices $i$, $j$, $k$, $l$, we use clauses $\lnot e_{ij} \lor \lnot e_{jk} \lor \lnot e_{kl} \lor \lnot e_{li}$, $\lnot e_{ij} \lor \lnot e_{jl} \lor \lnot e_{lk} \lor \lnot e_{ki}$, and $\lnot e_{il} \lor \lnot e_{lj} \lor \lnot e_{jk} \lor \lnot e_{ki}$ to encode the fact that a solution produced by the solver must be squarefree. By enumerating all possible choices of four vertices and constructing the above CNF formula, we force the graph to be squarefree.
The total number of clauses used is $3\cdot\binom{n}{4}$.

\subsection{Encoding the Minimum Degree Constraint}

For each vertex $i$, to ensure that $i$ is connected to at least three other vertices, we take each subset $S$ of $\{1,\dotsc,i-1,i+1,\dotsc,n\}$ with cardinality $n-3$ and construct the clause
$\bigvee_{j\in S} e_{ij}$.
By enumerating over all such subsets we enforce a minimum degree of~$3$ on vertex~$i$. Thus, constructing similar formulae for all vertices $1\leq i\leq n$, enforces that any vertex in the graph has a degree of at least $3$.
The total number of clauses used is therefore $n\cdot\binom{n-1}{n-3}=n\cdot\binom{n-1}{2}$.

\subsection{Encoding the Triangle Constraint}

We encode the property that every vertex is part of a triangle as follows: 
for each vertex $i$, we require $2$ other distinct vertices to form a triangle, and there are $\binom{n-1}{2}$ possible triangles containing $i$.
At least one of those triangles must be present in the
KS graph---this is encoded by the clause 
$\bigvee_{j, k \in S} t_{ijk}$
where $S$ is $\{1,\dotsc,i-1,i+1,\dotsc,n\}$ and $j<k$.
Using this clause for each $1\leq i\leq n$ ensures that every vertex is part of a triangle and hence there are $n$ triangle clauses.

\subsection{Encoding the Noncolorability Constraint}\label{color}
Recall that the key property of a KS graph is that it is non-010-colorable. As stated earlier, a graph is non-010-colorable if and only if for all $\{0,1\}$-colorings of the graph, a pair of color-1 vertices is connected or a set of three color-0 vertices are mutually connected.

For each $\{0,1\}$-coloring, a KS graph has a set~$V_0$ of color-0 vertices and a set~$V_1$ of color-1 vertices. Given a specific such coloring, the clause
\[ \bigvee_{\substack{i, j \in V_1\\i<j}}e_{ij} \lor \bigvee_{\substack{i, j, k \in V_0\\i<j<k}}t_{ijk} \]
encodes that this coloring is not a 010-coloring of a graph---since either a pair of color\nobreakdash-1 vertices is connected or three color-0 vertices are mutually connected. Note that we have to generate such a clause for all possible colorings, and conjunct them together to obtain a non-colorability constraint for graphs of order $n$. An assignment that satisfies such a constraint corresponds to a graph that is not 010-colorable under any possible coloring. Observe that in order $n$ the total number of such clauses is $2^n$. 

Fortunately, an empirical observation allows cutting the size of the formula dramatically: $\{0,1\}$-colorings with more than $\lceil\frac{n}{2}\rceil$ color-1 vertices are unlikely to be 010-colourings and in practice are not useful in blocking 010-colourable graphs.
Put differently, by dropping the constraints with $\lvert V_1\rvert\geq\lceil\frac{n}{2}\rceil$ we reduce the formula size drastically (making the formula easier to solve) and the corresponding increase in the number of satisfying assignments is small enough that these candidates can be ruled out via post-processing (Section~\ref{robust}).
In fact, for graphs up to order 23, no additional satisfying assignments (or candidate KS graphs) were generated. 


\subsection{Encoding Static Isomorphism Blocking Clauses}\label{sec:symbreaking}

Following~\cite{Codish2018}, we use symmetry breaking constraints that enforce a lexicographical order among rows of the graph's adjacency matrix. These small number of additional constraints enable us to {\it statically block} many isomorphic graphs. 

Given an adjacency matrix~$A$ of a graph,
we define~$A_{i,j}$ as the $i$th row of $A$ without columns~$i$ and~$j$.
Codish~et~al.~prove that up to isomorphism every graph can be represented by an adjacency matrix $A$ for which~$A_{i,j}$
is lexicographically equal or smaller than~$A_{j,i}$ for all $1\leq i<j\leq n$.

We express that $A_{i,j} = [x_1, x_2, \dotsc, x_n]$ is lexicographically equal or less than $A_{j,i} = [y_1, y_2, \dotsc, y_n]$
using $3n-2$ clauses and auxiliary variables $a_1$, $\dotsc$, $a_{n-1}$~\cite{knuth2015art}.
The clauses are $\lnot x_k \lor y_k \lor \lnot a_{k-1}$, $\lnot x_k \lor a_k \lor \lnot a_{k-1}$, and $y_k \lor a_k \lor \lnot a_{k-1}$ for $k = 1$, $\dotsc$, $n-1$.  The literal $\lnot a_0$ is omitted and the clause $\lnot x_n \lor y_n \lor \lnot a_{n-1}$ is also included.

\section{Orderly Generation via SAT+CAS}\label{orderly}

The symmetry breaking constraints described in Section~\ref{sec:symbreaking} do not block all isomorphic copies of adjacency matrices.
Thus, a crucial part of the \PC pipeline is the use of a SAT+CAS combination of a SAT solver and an
isomorph-free generation routine (the CAS part).
The orderly isomorph-free generation approach was developed independently by Read et al.~\cite{read1978every} and Faradvzev et al.~\cite{faradvzev1978constructive}. It relies on the notion of a \emph{canonical representation} of an adjacency matrix.

\begin{definition}[{\bf Canonical Graph}]\label{def:canonical}
An adjacency matrix $M$ of a graph is canonical if every permutation of the graph's vertices produces a matrix lexicographically greater than or equal to $M$,
where the lexicographical order is defined by concatenating the above-diagonal entries of the columns of the adjacency matrix starting from the left.
\end{definition}

An \emph{intermediate} matrix of $A$ is a square upper-left submatrix of $A$.  If $A$ is of order~$n$
then its intermediate matrix of order $n-1$ is said to be its \emph{parent},
and $A$ is said to be a \emph{descendant} of its intermediate matrices.

The OG method is based on the following two consequences of Definition~\ref{def:canonical}:
\begin{enumerate}[label={(\arabic*)}]
    \item Every isomorphic class of graphs only has exactly one canonical representative.
    \item If a matrix is canonical, then its parent is also canonical.
\end{enumerate}

Note that the contrapositive of the second property implies that if a matrix is not canonical, then all of its descendants are not canonical.
The OG process only generates canonical matrices and they are built starting from the upper-left.
Therefore, any noncanonical intermediate matrix that is encountered during an OG exhaustive search can be discarded, as none of its descendants will be canonical.

As described in Figure~\ref{fig:pipeline}, in our SAT+CAS implementation, when the SAT solver finds an intermediate matrix the canonicity of this matrix is determined
by a canonicity-checking routine implemented in the \PC system.
If the matrix is noncanonical, then a ``blocking'' clause is learned which removes this matrix (and all of its descendants) from the search.
Otherwise, the matrix may be canonical and the SAT solver proceeds as normal.

When a matrix is noncanonical, the canonicity-checking routine also provides a ``witness''
of this fact (a permutation of the vertices that produces a lex-smaller adjacency matrix).
We combine this process with the symmetry breaking clauses of Codish~et~al.~that canonical matrices can be shown to satisfy~\cite[Def.~8]{Codish2018}.

The OG technique provides a speedup that
seems to increase exponentially in the order $n$ of the KS graph---see Table~\ref{tbl:results}, which
provides experimental running times comparing
the SAT+CAS approach against SAT-only and CAS-only
approaches.  These timings were run on an Intel Xeon E5-2667 CPU and the CAS compared against was the \emph{nauty} graph generator~\cite{MCKAY201494}
with the same configuration from~\cite{uijlen2016kochen}. More details on our experimental setup can be found in Section \ref{results}. 

As described in Figure~\ref{fig:pipeline}, we simplify the SAT instance using the SAT solver
CaDiCaL~\cite{cadical} before solving the instance using MapleSAT~\cite{DBLP:conf/sat/LiangGPC16maple}.

\section{Embeddability Checking}\label{sec:embeddability}

\begin{table}
\begin{minipage}{200pt}
\begin{tabular}{cccc}
Order & Squarefree + Min.~Degree~2 & Min.~Unembed. & Runtime \\ \hline
4--9 & \phantom{00,}164 & \phantom{0}0 & 30 s \\
10 & \phantom{00,}563  & \phantom02 & 4.1 m\\
11 & \phantom03,257  & \phantom05 & 1.3 h\\
12 & 23,699 & 10 & 27 h\\
\end{tabular}
\end{minipage}
\caption{Counts for the number of minimal unembeddable graphs in orders up to 12 and the computation time for the embeddability check.}\label{tbl:embed}%
\end{table}

We refer to the solutions generated by the SAT solver as \emph{KS candidates}. Note that we have to additionally check whether a KS candidate is embeddable in order to detect whether it is a KS graph (and hence corresponds to a KS vector system). Hence, we perform an embeddability check on every KS candidate generated by the SAT+CAS solver of \PC.

Operationally, a graph $G$ is said to be embeddable if every pair of adjacent $G$-vertices can be mapped to two orthogonal vectors on the unit sphere in $\mathbf{R}^3$ (refer to Definition~\ref{embeddable}).
Otherwise, we say that $G$ is unembeddable. 

Our embeddability checking algorithm consists of two parts.
The first part is an integration of the vector assignment algorithm of~\cite{uijlen2016kochen} that finds all possible vector assignments describing the orthogonal relations between the vectors $v_i$ in a KS candidate defined by a set of edges $E$.
A vector assignment is a set of edge pairs $C=\{ (e_{ij}, e_{ik}), (e_{lm}, e_{ln}), \dotsc\}\subseteq E^2$ where each pair of edges share one common vertex and each pair is disjoint from each other, meaning the same edge cannot exist in more than one pair.
Each pair $(e_{ij}, e_{ik})$ in $C$ can be interpreted as a cross product relationship between the vectors $v_i$, $v_j$, and $v_k$, since the presence of $e_{ij}$ and $e_{ik}$ in the KS graph means that vector $v_i$ must be orthogonal to both $v_j$ and $v_k$ in any embedding of the candidate.

\begin{figure}[!ht]
    \centering
    \subfloat[\centering ]{{\includegraphics[width=6cm]{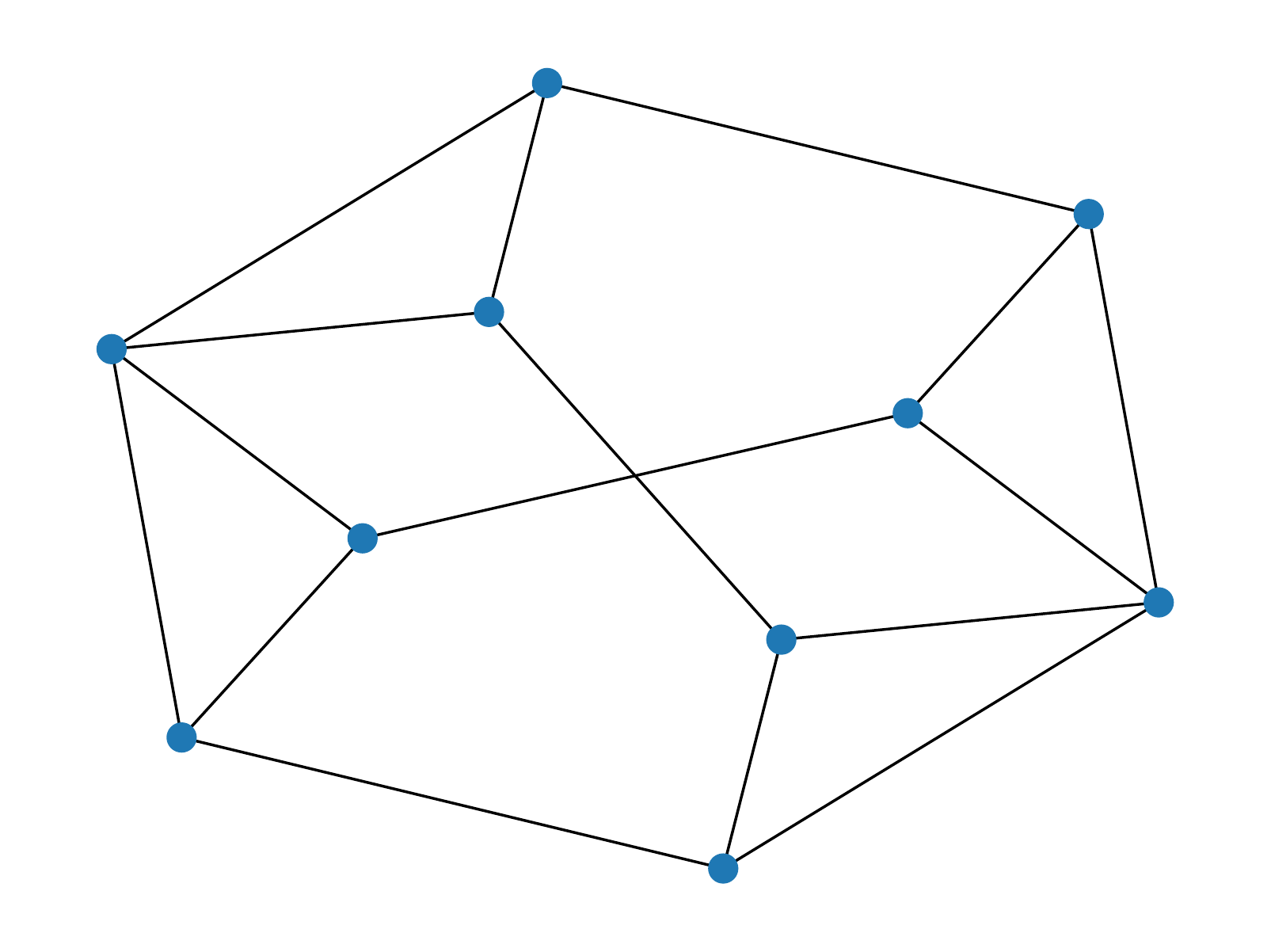} }}%
    \qquad
    \subfloat[\centering ]{{\includegraphics[width=6cm]{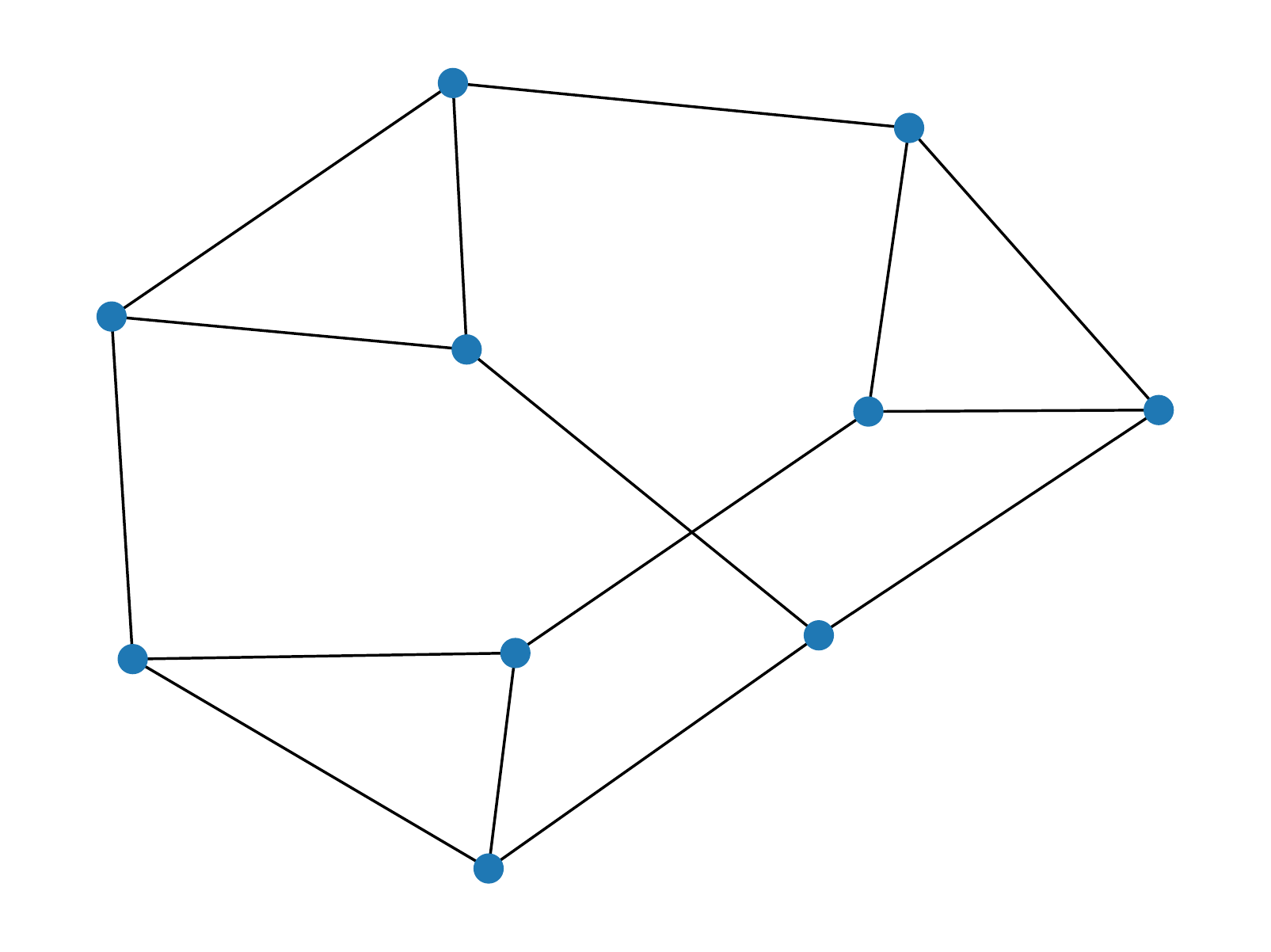} }}%
    \caption{The only two minimal nonembeddable graphs of order 10.  These are the smallest squarefree graphs that are not embeddable.}
    \label{fig:minimal}%
\end{figure}

The second part of the algorithm applies an SMT solver to determine the satisfiability of a system of nonlinear equations generated from a particular vector assignment as described below. More precisely, an assignment generated by Uijlen and Westerbaan's algorithm is converted into a set of cross and dot product equations, and these equations are passed to the theorem prover Z3~\cite{moura2008z3} that solves the equations over the real numbers.
We denote the vector corresponding to vertex $v_i$ as $V_i$ in Z3, where $V_i$ is a 3-tuple of real numbers.

Given a specific vector assignment generated by the previous algorithm, the system of constraints is as follows:

\begin{enumerate}
    \item If $(e_{ij}, e_{ik}) \in C$, we add the cross product constraint $V_i = V_j \times V_k$.
    \item If $e_{ij}$ is one of the edges of $E$ that is not contained in any of the pairs of $C$, we add the dot product constraint $V_i \cdot V_j = 0$.
    \item If $i \neq j$ then $V_i$ must not be collinear with $V_j$, so we add the noncollinearity constraint $V_i \times V_j \neq \vec{0}$.
\end{enumerate}

A \emph{free vector} is one that has not been fixed as the cross product of two other vectors. Of all possible assignments, we first choose the one with the least number of free vectors, since in practice such an assignment is likely to be solved more quickly.

It is important to note that for any vector assignment, each edge of the KS graph is encoded into either constraint~1 or~2, making the encoding shorter and more efficient than the naive encoding.
Constraint 3 requires two vectors to be noncollinear rather than only being nonequal since we do not enforce vectors to have unit length for reasons of efficiency.
This is a harmless optimization since vectors can be projected onto the unit sphere without disturbing these constraints.  

We also fix two orthogonal vectors to be the standard vectors $(1,0,0)$ and $(0,1,0)$ to cut down on the number of free variables.
To check whether a graph is embeddable, we use Z3 to determine whether these nonlinear arithmetic constraints are satisfiable over the real numbers.  Z3 applies a CDCL-style algorithm to decide the satisfiability of such systems~\cite{jovanovic2012solving}. 
If a solution is found, it is an assignment of vertices to vectors that satisfies all orthogonality constraints and the graph is therefore embeddable.

Embeddability checking of
large graphs can be further optimized by precomputing minimal unembeddable
graphs, as defined below.

\begin{definition}[{\bf Minimal Unembeddable Graph}]
An unembeddable graph $G$ is said to be a \textbf{minimal unembeddable graph} if any proper subgraph of $G$ is embeddable.
\end{definition}

A graph is unembeddable if and only if it contains a minimal unembeddable subgraph.
To optimize embeddability checking, we precomputed all minimal unembeddable graphs of orders up to and including~$12$.
We only consider squarefree graphs with a minimum degree of $2$ or greater, as the square graph $C_4$ is minimally unembeddable itself, and a graph containing a vertex of degree $0$ or $1$ is not minimally unembeddable.
We implement the unembeddable subgraph blocking technique as part of \PC.
If a graph contains one of the 17 minimal unembeddable graphs up to order 12 as a subgraph, the graph is blocked dynamically and is not considered as a satisfying assignment by \PC. This technique leads to a significant reduction of the number of satisfying assignment (candidates) generated by the SAT+CAS solver.

\section{Parallelization}\label{parallelization}

In prior applications of the CnC technique \cite{heule2017solving,heule2016solving}, the cubing solver generates a collection of cubes before the conquering solver is invoked.  Subsequently, each of these subproblems is solved using the conquering solver in parallel. However, this approach presents two primary challenges. Firstly, 
the generated cubes might exhibit imbalanced solving times, especially since the cubing solver does not have the ability to call the CAS to incorporate isomorph-free generation.
Secondly, the proof size for each subproblem also varies, making it difficult to allocate an appropriate amount of memory to individual cores.
In \PC, we implement a slight modification of traditional CnC practices to resolve the above challenges. In addition, we use a new CnC solver \AMS~\cite{jha2024alphamaplesat} that provides significant speedup for the cubing process. 

In our proposed method, the cubing solver of \AMS operates on a much smaller CNF instance obtained by omitting all non-colorability constraints (Section~\ref{color}). This approach is substantiated by empirical evidence suggesting that the same set of cubes can be obtained even without the non-colorability constraints. \AMS iteratively generates cubes until the total number of cubes surpasses a predefined cutoff, which is based on available computational resources.
Following cube generation, each subproblem is processed by the simplifcation solver (CaDiCaL with OG), and then it is passed to the conquering solver (MapleSAT with OG) and solved in parallel. To efficiently manage the termination of each subproblem, a termination strategy is implemented: if the proof size for a subproblem exceeds 7~GiB, the problem is further divided into more cubes and solved accordingly. This iterative process continues until every subproblem can be resolved with proof sizes under 7~GiB. We overcome the challenges posed by varying proof sizes by implementing this slight modification and it allows us to verify all generated proof certificates with at most 4~GiB of memory allocation.
\section{Results}\label{results}

\subsection{Experimental Setup}\label{setup}

Results in Table~\ref{tbl:results} were conducted on a cluster of Intel E5-2683 v4 Broadwell @ 2.1GHz CPUs, each with access to 4 GiB of RAM and running 64-bit CentOS Linux 7. 
Results in Table~\ref{tbl:resultsp} were conducted on a cluster of Dual Intel Xeon Gold 6226 CPUs @ 2.7 GHz (24 cores/node). We used the \texttt{g++} compiler version 9.3.0 with option \texttt{-O3} to compile the SAT solvers used.
24 CPUs were used to solve and verify order 22 (in about 4 days wall clock time), and 240 CPUs were used for order 23 (in about 7 days wall clock time).
To compare \PC against other approaches, we computed the total CPU time used by all processes in the solving process, including simplification, cubing, SAT solving, CAS queries, and embeddability checking.

\subsection{Key Findings}

\begin{table}
\centering
\resizebox{0.7\textwidth}{!}{ 
\small 
\begin{tabular}{ccccc}
\toprule
$n$ & SAT+CAS & SMS & CAS-only & SAT-only \\
\midrule
17 & \phantom{000}0.3 m &  \phantom{000}0.2 m & \phantom{00}25.2 m & \phantom{,0000}9.0 m \\
18 & \phantom{000}1.8 m &  \phantom{000}1.2 m & \phantom{0}455.4 m & \phantom{,00}266.4 m \\
19 & \phantom{000}9.0 m &  \phantom{000}8.4 m & 9506.4 m & 11,705.8 m \\
20 & \phantom{0}140.5 m &  \phantom{0}100.8 m & \phantom{0}timeout & \phantom{,00}timeout \\
21 & 1945.0 m & 1574.4 m & \phantom{0}timeout & \phantom{,00}timeout \\
\bottomrule
\end{tabular}
}
\caption{SAT+CAS vs.\ SMS, CAS-only (nauty), and SAT-only: The total CPU time (in minutes) compared on orders $17\leq n\leq 21$. All tools are sequential. For higher orders, CAS-only and SAT-only time out at 12,000 minutes.}
\label{tbl:results}
\end{table}

\begin{table}
\centering
\resizebox{0.7\textwidth}{!}{ 
\small 
\begin{tabular}{c@{}cccc@{\quad}cc}
\toprule
&& \multicolumn{2}{c}{Number of Cubes} & & \multicolumn{2}{c}{Total CPU time} \\
$n$ && SAT+CAS & SMS & & SAT+CAS & SMS \\
\midrule
22 && \phantom{0}26,646 & \phantom{0}18,659 && \phantom{,00}932 h & \phantom{,00}628.1 h \\
23 && 173,097 & 313,665 && 12,116 h & 11,921.8 h \\
\bottomrule
\end{tabular}
}
\caption{Parallel CnC SAT+CAS vs.~parallel SMS: The number of cubes and total CPU time for the parallel versions of SAT+CAS and the SAT Modulo Symmetries tool on orders $22\leq n\leq 23$.}\label{tbl:resultsp}
\end{table}

\begin{figure}[!ht]
    \centering
    \subfloat[\centering ]{{\includegraphics[width=6cm]{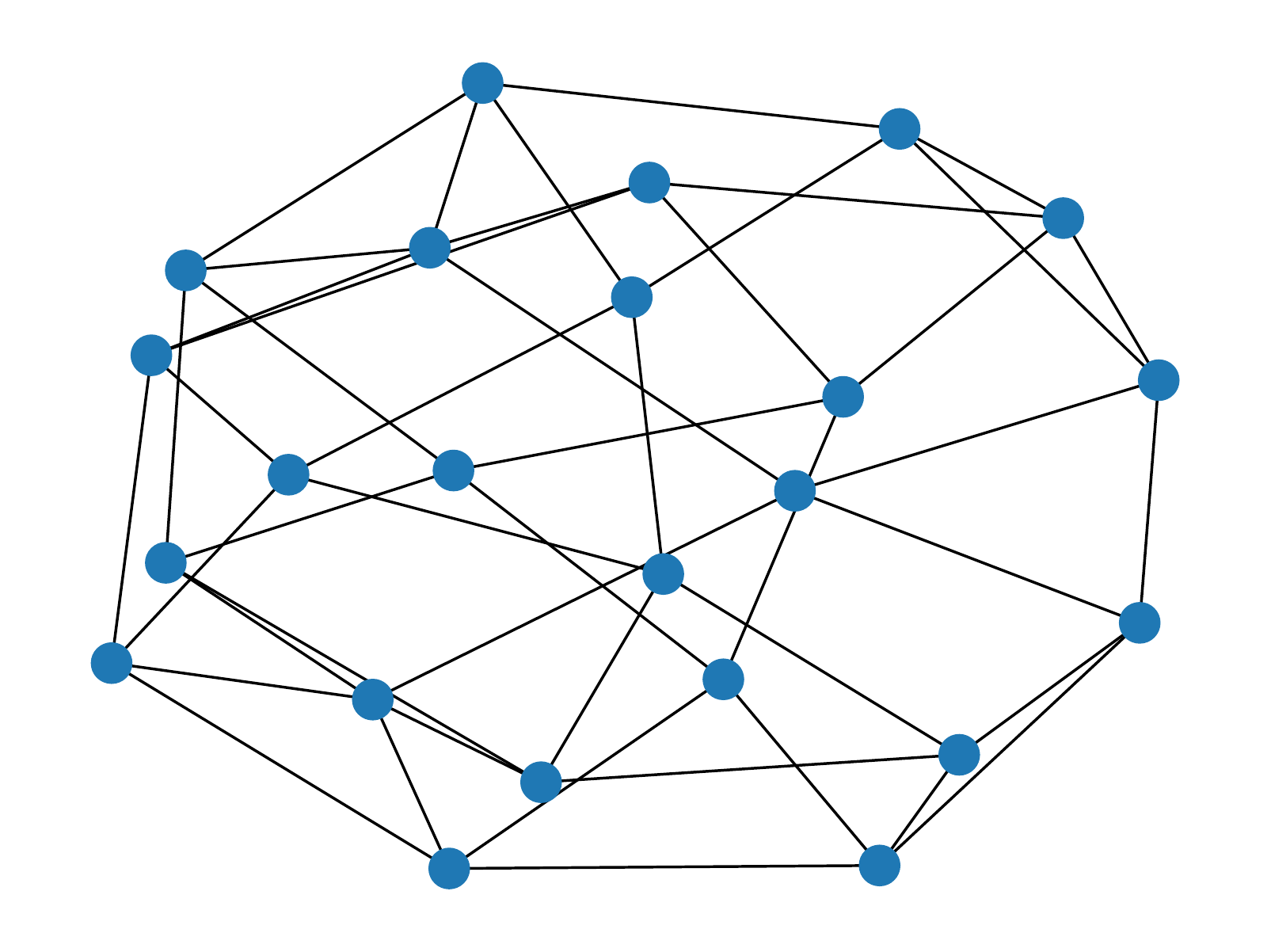} }}%
    \qquad
    \subfloat[\centering ]{{\includegraphics[width=6cm]{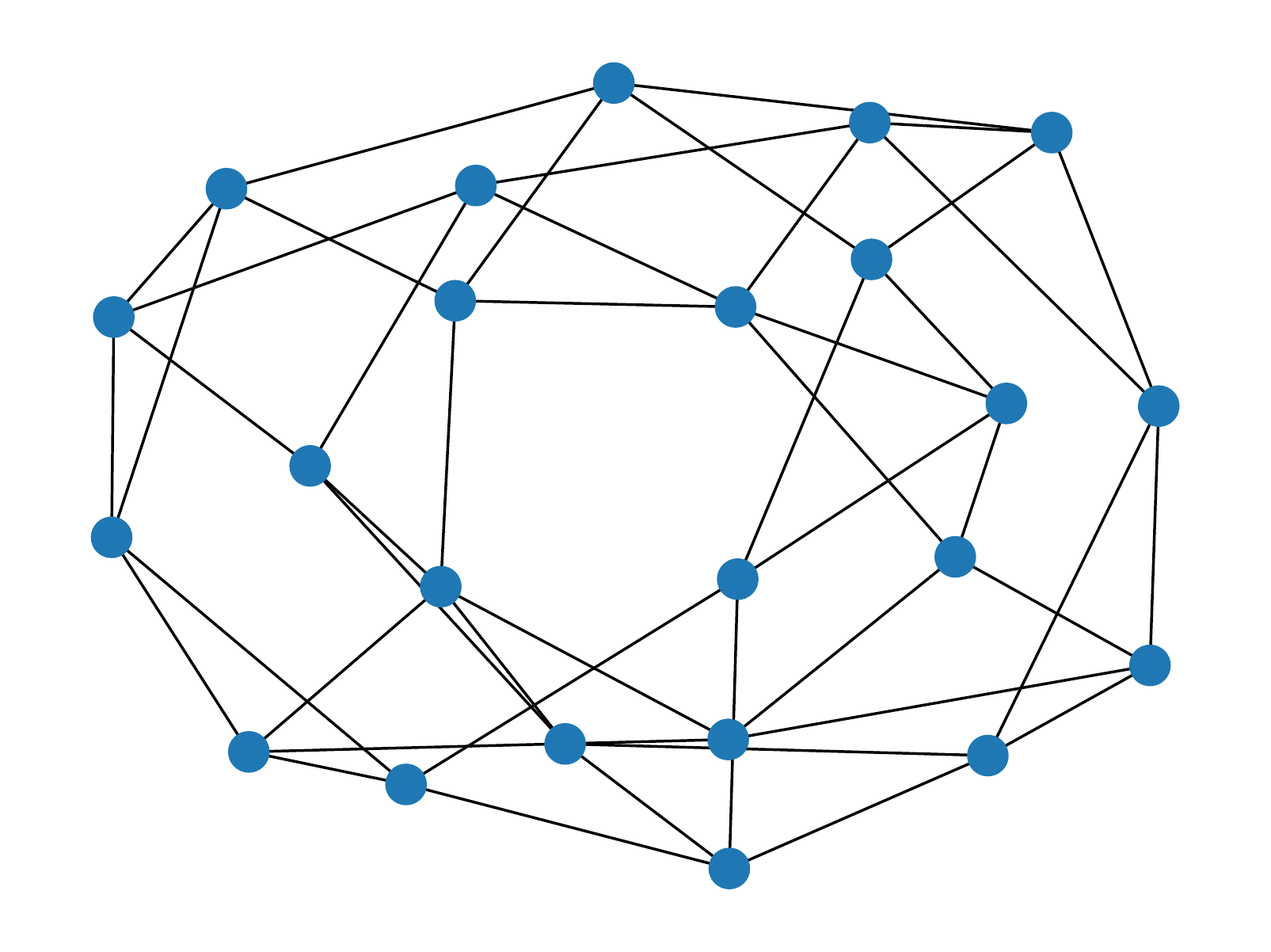} }}%
    \caption{The only two KS candidates up to order~23 that do not contain minimal nonembeddable graphs up to and including order 14. These two graphs are determined to be unembeddable by Z3.}
    \label{fig:23}%
\end{figure}

Tables~\ref{tbl:results} and~\ref{tbl:resultsp} contain our experimental results. As can be seen from Table~\ref{tbl:results}, the sequential version of \PC outperforms the sequential SAT-only (MapleSAT) and CAS-only (nauty) tools. Further, the sequential \PC is comparable to sequential SMS on orders 17 to 21, even with proof production switched on. (Note that we are using the results reported by the SMS authors~\cite{KirchwegerPeitlSzeider23} which did not include generating proofs.)

The sequential \PC~also outperforms the parallel tool of~\cite{uijlen2016kochen}.
For example, for order 21, sequential \PC takes about 32.4 hours
while their tool took 3 months on 300 CPUs in 2014 (the exact kind of CPUs used are not specified).
Further, their tool does not solve orders 22 or greater.

For higher orders of KS (i.e., 22 and 23), we do not expect the sequential versions of \PC and SMS to scale in a reasonable amount of time, and hence we compare only their parallel CnC versions (Table~\ref{tbl:resultsp}). The authors of SMS also do not report on the performance of their sequential version for orders 22 and 23. 
In comparison to SMS, our approach was hindered due to our generating and verifying all proof certificates---requiring us to ensure each proof certificate did not grow too large in order to permit effective verification.
Such an approach requires splitting each cube whose proof certificate grows too large, and this additional splitting slows down our approach. 

Our method benefits from dynamically blocking minimal unembeddable subgraphs up to order 12, which enables \PC to generate significantly fewer KS candidates compared to \cite{KirchwegerPeitlSzeider23}. Specifically, we generate a single candidate in order 22 and 41 candidates in order 23, while \citeauthor{KirchwegerPeitlSzeider23}\ generates 88,282 and 3,747,950 candidates respectively.

\begin{figure}[!ht]
\centering
\includegraphics[scale=0.5]{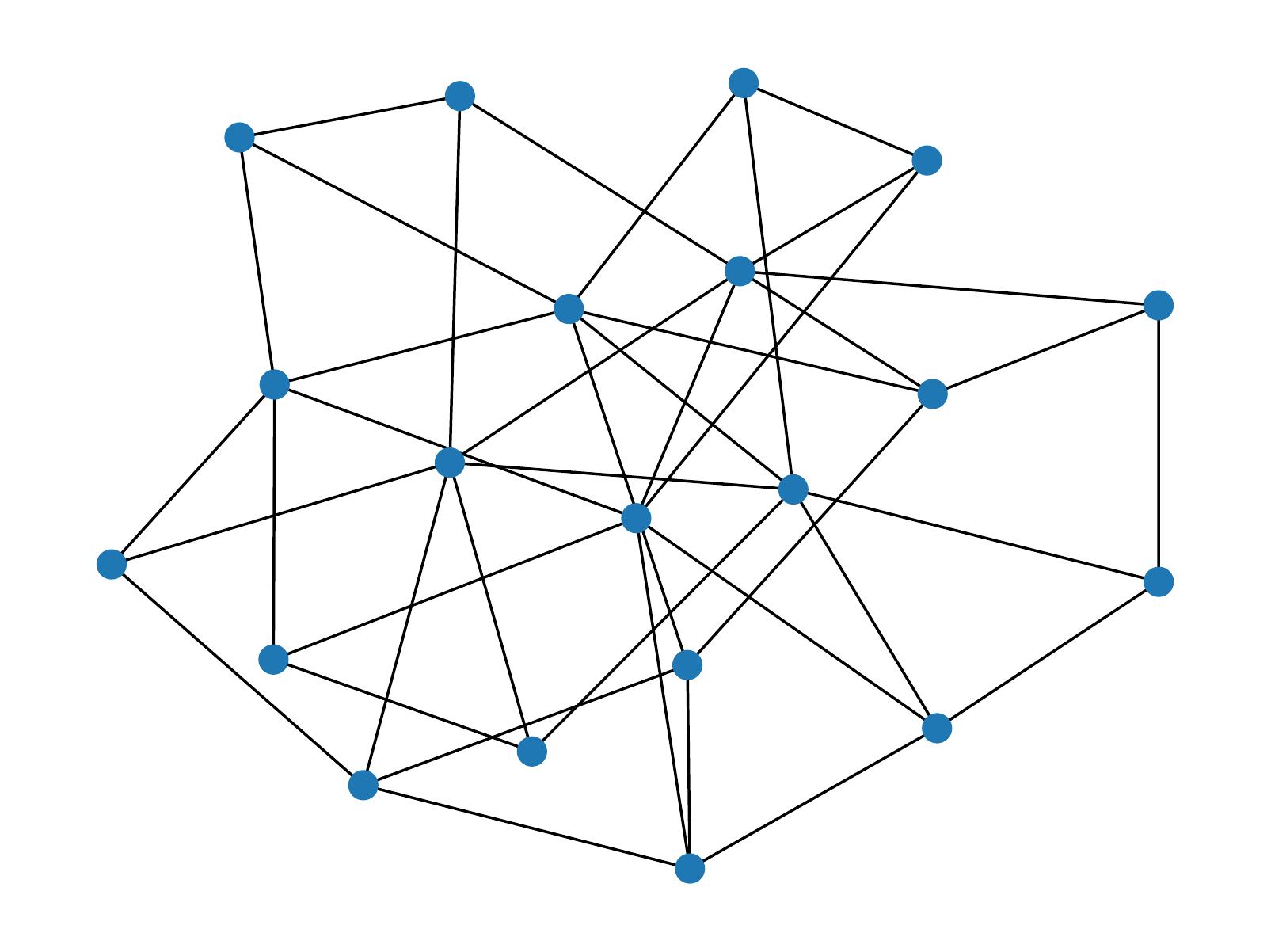}
\caption{One of the four graphs with 20 vertices that were not present in Uijlen and Westerbaan's enumeration. The four graphs satisfy all constraints mentioned in Section~\ref{sec:encoding}, but are not embeddable, and therefore do not constitute a KS system. }
    \label{fig:order20}
\end{figure}

We compared our KS candidates with Uijlen and Westerbaan's
findings, and verified their conclusion that there is no KS system with strictly less than 22 vectors. In order $20$, we found four additional KS candidates that were not present in the collection of Uijlen and Westerbaan, indicating that their search missed some KS candidates. We present one of the missing graphs in Figure~\ref{fig:order20}. We verified that these four additional graphs satisfy the constraints of a KS candidate
and therefore would be KS systems were they embeddable, but unfortunately, they are not.

Another notable question we address is the existence of a complex KS vector system. Specifically, is there a KS graph that can be embedded over the complex sphere? We employ a similar embeddability verification pipeline as mentioned earlier, with the distinction that vectors now may possess complex number coordinates. We discover that each KS graph up to order 23 either contains a complex unembeddable subgraph or is determined to be unembeddable over the complex numbers via Z3.
Therefore, we conclude that the minimum size for both the real and complex KS system are at least $24$.

\section{Verification of Results}\label{robust}

In order to verify the computations produced by \PC,
we use DRAT proof logging which is standard in all modern SAT solvers.
This makes it possible for a third-party proof verifier to provide an
independent certification of the correctness of the solver's conclusion
(assuming the input Boolean formula correctly encodes the minimum KS problem).
A DRAT proof consists of a trace of the clauses learned by the solver during its execution.
A proof verifier checks that each clause can be derived from the previous clauses using simple rules known to be logically consistent.
The CAS-derived noncanonical blocking clauses cannot be verified using the normal rules,
so we specially tag them to be verified separately.
Specifically, they are justified via a CAS-derived permutation that,
when applied to the blocked adjacency matrix, produces a
lex-smaller adjacency matrix---and therefore provides a witness that the blocked matrix is noncanonical and is safe to block.
Similarly, the unembeddable subgraph blocking clauses are justified by a CAS-derived
permutation that when applied to the blocked subgraph transforms it into one of the
minimal unembeddable subgraphs (see Section~\ref{sec:embeddability}).

The CAS-derived clauses in the DRAT proof are prefixed by the character `\texttt{t}' to signify they should be trusted
and we modified DRAT-trim~\cite{wetzler2014drat} to trust such clauses
following the approach first used in~\cite{Bright2020}.
The trusted CAS-derived clauses are separately verified by
a permutation-applying Python script that applies the witnesses
produced by the CAS to verify the blocked matrices are noncanonical or unembeddable.
Similarly, when a KS candidate is found, the solver learns a trusted clause blocking the candidate
(so that the search continues until all candidates have been found).
The DRAT proof ends with the empty clause, which by definition is not satisfiable. If the verifier is indeed able to verify the empty clause then we can have confidence that the SAT solver's search missed no candidates without needing to trust the solver.

All results have been certified.
The uncompressed proofs for order 22 are about \proofsizetwo~TiB in
total, and \proofsizethree~TiB for order 23. The certification for orders~22 and~23 required using CnC (as described in Sec.~\ref{parallelization})
to ensure that each DRAT proof could be verified with at most 4~GiB of memory.

All KS candidates produced by our method have been extensively checked.
For example, each KS candidate is passed into a verification script implemented using the NetworkX~\cite{SciPyProceedings_11} graph package to verify that they satisfy all encoded constraints (see Section~\ref{sec:encoding}). Further, we test the embeddability pipeline by performing a verification on all 
squarefree graphs of order up to 12 with minimal vertex degree 2.
Specifically, if a graph is embeddable and corresponds to a set of vectors, we check that no pair of vectors in the set are collinear, and a pair of vectors are orthogonal if their corresponding vertices are connected. 

\section{Conclusion}\label{conclusion}

We give a computer-assisted proof showing that a KS vector system in three dimensions must contain at least 24 vectors. Crucially, our proof is verifiable by an independent third-party proof checker.
In addition, we provide a computational speedup of over four orders of magnitude over the previously used approach of Uijlen and Westerbaan~\cite{uijlen2016kochen}.
For the first time, we successfully implemented and applied the SAT+CAS paradigm along with orderly isomorph-free generation to provide a robust pipeline for problems in quantum foundations. The validity of our work is further confirmed by Kirchweger, Peitl, Szeider~\cite{KirchwegerPeitlSzeider23}, who performed an independent search for KS~systems with up to 23 vectors with a similar approach as our technical report~\cite{tech} using a SAT modulo symmetries (SMS) solver.
Compared to previous work, our approach is less error-prone since we use heavily-tested proof-generating SAT solvers such as MapleSAT and CaDiCaL.
We verified the produced proofs using independent proof checkers, meaning our result does not rely on the correctness of MapleSAT or CaDiCaL.

Finding the minimum KS system has remained stubbornly open for over 55 years. It is not only a problem of great importance to quantum foundations, but has direct applications to various fields of quantum information processing, such as quantum cryptographic protocols~\cite{cabello2011hybrid}, zero-error classical communication~\cite{cubitt2010improving}, and dimension witnessing~\cite{guhne2014bounding}. As a consequence, a wide variety of techniques have been developed to address this question over the past several decades. We add a novel class of techniques to this body of work.

The SAT+CAS paradigm has been successfully used to resolve a number of mathematical problems in combinatorics, number theory, and geometry that had previously remained unsolved for many decades~\cite{bright2016mathcheck, Heule2021, bright2021sat}. With this work, we extend the reach of the SAT+CAS paradigm, for the first time, to resolving combinatorial questions in the realm of quantum foundations.

\section*{Code Availability}
The \PC~pipeline is open-source and can be accessed at \url{https://github.com/BrianLi009/MathCheck}.

\section*{Acknowledgements}
This research was enabled in part by support provided by Compute Ontario (\href{https://www.computeontario.ca/}{https://www.computeontario.ca/}) and the Digital Research Alliance of Canada (\href{alliancecan.ca}{alliancecan.ca}).

\bibliography{sn-bibliography}

\end{document}